\documentstyle[aps,preprint]{revtex} 
\newcommand{\U}{{\bf u}}

\newcommand{\X}{{\bf x}}

\tightenlines 
\begin{document}

\title{Fractal rain distributions and chaotic advection} 
 
\author{Ronald Dickman$^\dagger$ 
} 
\address{ 
Departamento de F\'{\i}sica, ICEx, 
Universidade Federal de Minas Gerais,\\
Caixa Postal 702, 
30161-970 Belo Horizonte, Minas Gerais, Brazil} 
\date{\today} 
 
\maketitle 
\begin{abstract} 
Localized rain events have been found to follow power-law
distributions over several decades, 
suggesting parallels between precipitation and seismic
activity [O. Peters et al., PRL {\bf 88}, 018701 (2002)].
Similar power laws can be generated by treating 
raindrops as passive tracers advected by the velocity field
of a two-dimensional system of point vortices
[R. Dickman, PRL {\bf 90}, 108701 (2003)].
Here I review observational and theoretical aspects of
fractal rain distributions and 
chaotic advection, and present new results on tracer
distributions in the vortex model.
\vspace{2em}  

\noindent PACS numbers: 89.75.Da, 47.27.Eq, 92.40.Ea, 92.60.Ek 
\vspace{2em}

\noindent $^\dagger${\small Email address: dickman@fisica.ufmg.br}

\end{abstract} 

\newpage

\section{Introduction}

Complex systems often exhibit fractal or power-law
scaling; Earth's atmosphere is
no exception.  Fractal rain distributions have
been known for at least two decades \cite{lovejoy85,olsson,lavergnat},
while recent analyses indicate that durations of
dry intervals, and the size of rain events,
follow power laws \cite{peters,christensen}.
The similarity between the latter observations and scaling laws in
seismic activity suggests a parallel between
rain and earthquakes, and a possible connection
with the phenomenon of self-organized criticality \cite{christensen,btw}.  

Atmospheric motion is turbulent,
particularly in the vicinity of storms, and various aspects
of turbulent flow follow power laws 
over many orders of magnitude \cite{mccomb,frisch,lovejoy}.
Even in the absence of fully developed turbulence, 
unsteady flow may stretch and fold an
initially compact region, leading to a highly convoluted,
nonuniform density of suspended particles or droplets 
\cite{tritton,grebogi,elperin} via chaotic advection 
\cite{aref84,leoncini}.
In light of these observations, it is interesting to develop a
model in which rain is an {\it ideal passive tracer}
\cite{mccomb12,falkovich}. 
In \cite{rain} it was shown that such a model is capable of producing 
power-law-distributed event sizes and durations.

In this paper I review some of the evidence for
fractal rain distributions, and present new results on the
spatial distribution of tracers in the vortex model.  
Progress in fluid mechanics depends heavily on numerical solution
of the equations of motion, which in turn represents one of the 
most challenging areas in computational physics,
the theme of the present number.

In Sec. II, I survey observations of fractal rain distributions.
Sec. III contains a brief discussion of SOC-like approaches, while Sec. IV 
reviews results on tracer-particle dynamics in a fluid
undergoing chaotic advection. I define the vortex model
in Sec. V, which also includes a summary of previous findings and
some recent extensions.
Sec. VI presents new results 
on spatial distributions of tracers in the vortex model.  The paper closes in
Sec. VII with a summary and discussion of open questions.
 
\section{Fractal rain distributions}

Discussions of fractal rain distributions go back at least to
the work of Lovejoy and Mandelbrot \cite{lovejoy85} who presented
a model with a single fractal dimension.  The distribution in
question involves a time series of duration $T$ and a fixed observation
point or {\it station}.  The observation interval is partitioned
into $N = T/\tau$ subintervals of duration $\tau$, each
characterized as rainy (a nonzero amount of rain
is detected at the station in this interval) or dry.  
The function $r(\tau)$ is then defined as the number 
of rainy subintervals at scale $\tau$.  Olsson et al. found
that this distribution follows a power law, $r \sim \tau^{-\gamma}$,
with $\gamma \simeq 0.8$, over a certain range of durations \cite{olsson}.  
Note that for $\tau \approx T$,
$r \to N$ (all subintervals are rainy), while for $\tau$ much shorter than
the characteristic time between raindrops, $r$ saturates at a value $M$
equal to the total
number of raindrops incident on the station during the interval $T$.
Between these simple limits, $r(\tau)$ may exhibit nontrivial
behavior reflecting correlations in the generation or dynamics of raindrops.
Now, if the arrival times of the raindrops were mutually independent
(so that the time interval between successive drops at the detector
were exponentially distributed), the number of drops $n(\tau)$ in a
given subinterval would be Poisson-distributed with mean 
$\langle n(\tau) \rangle = m\tau$, with $m = M/T$, and we would have
$r(\tau) = (T/\tau)(1 - e^{-m\tau})$.   Thus a power law
distribution with $\gamma < 1$ rules out a simple 
``independent event" model, 
suggesting some nonlinear mechanism behind
the observed rainfall statistics.

The observations of Olsson et al. (from Sweden) were later corroborated 
by Lavergnat and Gol\'e \cite{lavergnat} in an experiment performed near
Paris.  The latter study generated data on raindrop arrival
times and sizes over a 14-month period, and confirmed
the scaling $r \sim \tau^{-0.82}$ over six orders of magnitude
(from 0.01 to 10$^4$ minutes).  Other important conclusions from this
study are: (1) the raindrop {\it diameter} distribution decays roughly 
exponentially (or perhaps as a stretched exponential) for diameters
greater than about 0.5 mm; (2) the distribution of time intervals
between raindrops can be fit to a 
so-called bi-Pareto distribution
over about nine orders of magnitude.  This distribution involves two power law
regimes, one for short times (drops associated with a given storm)
another for long times (intervals between successive storms).  
On the basis of their analysis Lavergnat and
Gol\'e conclude that the waiting time $D$ between successive rain events
is power-law distributed: $P_d(D) \sim D^{-\tau_D}$ with 
$\tau_D = 1.68$.  (For $D \approx$ one day the probability density
$P_d$ decays rapidly; droughts longer than a week or
so were not seen in their experiment.)

Convincing evidence for a multifractal {\it spatial} distribution of
raindrops in storms, on scale from 1 cm up to meters, was very recently
reported by Lovejoy et al. \cite{lovejoy03}.  An important conclusion
of these authors is that there is no meaningful way to describe rain
content in the atmosphere in terms of a smoothly varying density,
since large fluctuations are present at all scales.  The authors suggest
turbulence as the reason for the fluctuations in raindrop
distribution.

Recently a large time-series (six months) from radar 
observations on the
Baltic coast became available under the BALTEX project \cite{baltex}.  
The radar station
determines the quantity of rain falling
in a 1 m$^2$ column of the atmosphere.  Arrival times of
individual raindrops are not resolved, but the total amount
of rain above the station at each 1 min. interval is registered.
The threshold for detection is 0.005 mm/h; intervals with a precipitation
rate above this threshold have a nonzero rate $q(t)$, otherwise
$q(t) = 0$ for that interval.  In their analysis of the BALTEX data,
Peters et al. focus on {\it rain events}, defined as sequences of
consecutive intervals with nonzero rainfall \cite{peters,christensen}.  
A series of consecutive intervals having zero rain defines a drought.
The intensity $I = \sum_t q(t)$ of a rain event is the rainfall integrated over its
duration.  Peters, Hertlein and Christensen
found that the distribution of rain-event sizes at the Baltic coast
station follows a power law over at least three decades. Drought durations
are also power-law distributed over the 
range of several minutes to about a week, with a significant 
perturbation apparently reflecting diurnal variation.  The power laws identified
by Peters et al. may be expressed in the form
\begin{equation}
P_i(I) \sim I^{-\tau_I}
\label{PI}
\end{equation}
and 
\begin{equation}
P_d(D) \sim D^{-\tau_D}
\label{PD}
\end{equation}
where $P_i$ and $P_d$ are the probability distributions for rain event intensities,
and for drought durations, and the exponents
are found to take the values
\begin{equation}
\tau_I = 1.36 \;\;\;\;\;\;\;\; \tau_D = 1.42
\label{expP}
\end{equation}
These authors emphasize the similarities between these distributions and
those found for earthquakes, suggesting a parallel with self-organized
criticality to be discussed in the following section.

Taken as a whole, the observations of Olsson et al., Lavergnat and Gol\'e,
Lovejoy and co-workers, and Peters et al. present a very strong case
for fractal or multifractal distributions of rain at a given position over time, and in
space, at a given instant \cite{note1}.  The {\it universality} of the observed
distributions is less clear.  First, the time series all come from
the north of Western Europe, where prolonged dry periods are 
evidently rare.  The central region of Minas Gerais, Brazil (to cite
one example) experiences a dry spell of several months each year, and
might therefore exhibit a different distribution of droughts. 
The Paris and Sweden experiments yielded similar values ($\gamma = 0.82$) for
the exponent characterizing the fractal distribution in time,
while the BALTEX data yield $\gamma \simeq 0.55$ \cite{christensen}. 
On the other hand, the Paris results suggest
$\tau_D = 1.68$, considerably larger than the Baltic observations.
Observations from other sites (in particular, from other
regions of the world, including continental sites, and oceans),
are needed confirm the generality of power laws, and the range 
of exponent values.

\section{Rain and Self-organized criticality}

Peters, Hertlein and Christensen noted a striking similarity
between the scaling laws they found in the rain data and
those known for earthquakes.  Specifically, earthquake
magnitudes $M$ (defined in terms of energy released)
follow the Gutenberg-Richter law $P_m(M) \sim M^{-\tau_M}$ \cite{gutenberg},
while the waiting time between earthquakes in a given
region follows a power-law known as 
Omori's law \cite{omori,unif}. 
This suggests a parallel between
precipitation in the atmosphere and relaxation of the Earth's crust at
stressed tectonic-plate boundaries \cite{christensen}.  In the
context of seismology, cooperative relaxation due to elastic interactions
and nonlinear friction is captured by block-spring models 
\cite{burridge,carlson} or, in much-reduced fashion, by
sandpile models \cite{btw}.  The latter have attracted much attention  
as the principal example of the self-organized criticality paradigm
for scale-invariance in natural, far-from-equilibrium systems 
\cite{btw,socbjp,frigg}.

Indeed, sandpile-like models of rainfall have been 
studied \cite{pinho,andrade}.  
They involve the directed motion of raindrops 
such that when a given cell contains more than a certain number of
drops, the latter move to cells at the level below.
That such a model yields power-law distributions for sizes
of certain kinds of events
is not surprising, as this is an intrinsic feature of sandpile
models \cite{socbjp,note2}.   (It is less clear how to 
define the {\it duration} 
of a rain event, since sandpiles represent a singular limit in which
event durations cannot be measured on the same time scale as intervals
between events \cite{ggrin,vespignani}.)

But if certain aspects of rain distributions resemble those of avalanches in
sandpile-like models, the underlying physics remains obscure.
While it may yet prove possible to explain the observed power laws
in terms of an open, driven dissipative system \cite{pinho,andrade,lu},
there is no obvious reason for the formation or precipitation 
of one raindrop to provoke similar events nearby.  
Given the attendant release of latent heat,
one might instead expect a self-limiting tendency in condensation.

In fact, condensation and precipitation of rain is a complex process,
involving the interplay between atmospheric motion, including 
turbulent convection, thermodynamics and nucleation processes \cite{manton}.
Evaporation, condensation and vertical fluid motion are strongly
coupled via buoyancy.
While it is hard to see how direct interactions between raindrops over
a mean interparticle distance of 10 cm \cite{lovejoy03} could
lead to clustering, the drops are of course highly influenced by
the motion of the surrounding air.  The latter is {\it generically}
turbulent \cite{lovejoy}, and as such is characterized by scale-invariant
velocity and energy distributions.
Thus it appears more promising to seek the explanation for
power-law distributions in atmospheric fluid dynamics.

\section{Chaotic Advection}

In this and the following sections we will be interested in the
dynamics of {\it passive tracer particles} in a fluid.
Such a particle follows the local velocity of the fluid at each
moment, so that its trajectory is that of a fluid particle.
The fluid velocity is not affected by the tracers.  
As such, a tracer represents an idealized limiting case of a very small,
neutrally buoyant particle immersed in the fluid.  
(Tracers are small in the sense that (1) their inertia is negligible and
(2) the fluid velocity varies little over the diameter of the tracer.)
The idea of the
model to be developed below is that raindrops can be treated, to a first
approximation, as passive tracers, even though they are much denser
than air, and not always ``small."  This study should nevertheless
provide a preliminary indication of how
atmospheric motion can affect the distribution of the raindrops.

Now, if the fluid motion is turbulent, the distribution of passive
tracer particles should also exhibit scale-invariant properties
\cite{elperin,mccomb12,falkovich}.  An important 
example is Richardson's law, the empirical result that in turbulent flow,
the mean-square separation $\ell_t$ between a pair of tracers
at time $t$, given an initial separation of $\ell_0$, grows
$\sim \ell_0^{4/3}$.
If two or more tracers are released at nearby points, we can study
how their trajectories separate over time, leading to the notion
of chaotic tracer motion: trajectories that separate exponentially
rapidly with time.  A flow need not be turbulent to exhibit chaos in this
sense.  Relatively simple flows, such as the van Karman vortex street
or flows generated by systems of point vortices exhibit this property.  
Aref showed that this phenomenon, known as {\it chaotic advection}
or {\it Lagrangian chaos}, appears in systems
of as few as four mutually interacting vortices 
\cite{aref84,leoncini}.  (The vortex system, which is central to the
model developed here, will be described in detail below.)

Some aspects of chaotic advection can be 
understood in a general way using
elementary notions from dynamical systems theory.  Consider an
incompressible fluid restricted to a finite volume.  A stagnation
point in such a flow is a hyperbolic fixed point: due to volume
conservation, the fluid is attracted to this point along one
direction, and repelled along another.  As a result, a fluid element
that passes near the hyperbolic point is stretched along one
direction, compressed along the other.  As stretching continues, the
element must double back on itself since it is confined to a finite region.
Repeated encounters with hyperbolic points lead to iterated distortions of
the kind described above.  Thus a fluid element
undergoes repeated stretching and folding similar to the distortions 
leading to chaos in simple model systems such as the baker's 
transformation \cite{ottino}.

Flow fields with chaotic advection may also exhibit unstable
periodic orbits with fractal structure\cite{grebogi}; 
tracers (as well as particles with 
non-negligible inertia) may spend long periods of time 
in the vicinity of these orbits \cite{tel}.  The effect, once again, 
is that an initially compact region becomes highly extended along 
one direction, and contracted in the other, and repeatedly
folded, yielding a self-similar structure
of bands reminiscent of a strange attractor in a chaotic dynamical
system.   

Summarizing, the motion of tracer particles in even moderately
complex flows can
yield chaotic trajectories and scale-invariant spatial distributions.
This suggests treating rain as a collection of passive tracers 
moving in a chaotic or turbulent velocity field.
The raindrops are released in a localized condensation event,
and then advected by the air before being detected at or above a given point
on Earth's surface.

What would a reasonably complete model of this process look like?
Even ignoring thermodynamic aspects (evaporation and re-condensation of rain,
with attendant latent-heat and buoyancy effects), we would need to
treat a three-dimensional atmosphere whose density falls off exponentially
with height, and integrate the Navier-Stokes equation for an
incompressible fluid subject to suitable boundary and initial conditions,
(including a driving term at large scales to compensate small-scale
dissipation, if we wish to study a stationary state), at a Reynolds 
number characteristic of turbulent motion \cite{note3}.  
To include the possibility of convection we would need to implement
(at least) the Boussinesq approximation, allowing the density to vary
linearly with temperature, and including heat transfer in the
description \cite{chandra,manneville}.  Such a study poses a great 
challenge to presently avaliable
computational tools.  In particular, faithful representation
of fully developed turbulence appears (due to the number of degrees 
of freedom involved) computationally nonviable, so that reduced
descriptions such as large-eddy simulation or a shell model
are required \cite{mccomb,frisch}.  

While semi-realistic
simulation seems a worthy objective for future study, in this 
work I consider a radically simplified model, which can serve as
a proof of principle of the idea that fractal rain distributions
derive from chaotic advection.  The model eliminates nearly
all atmospheric processes and takes advantage of a physical
system (point vortices) affording a vast reduction in computational
complexity, as explained in the next section.

\section{Computational Model}

On the planetary scale, Earth's atmosphere is two-dimensional.
At high Reynolds numbers, effects of viscosity are limited to small
scales and to boundary layers.  These observations
may be seen as possible motivations for what is in the final
analysis a simplification based on computational necessity,
namely, the study of ideal two-dimensional flow.
In {\it potential flow}, i.e.,
for which the fluid velocity $\U (\X,t)$ can be written as the
gradient of a scalar function $\phi (\X,t)$ 
\cite{LandL,sommerfeld,note5},  
the incompressibility condition $\nabla \cdot \U = 0$
implies that $\phi$ satisfies Laplace's equation; such flows are
irrotational, i.e., $\nabla \times \U = 0$.  Potential flow
solutions of Euler's
equation satisfy the principle of linear superposition. 

The velocity field is built up out of complex potentials of the form
\begin{equation}
\phi = -i \frac{K}{2\pi} \ln (x + iy)
\label{vortpot}
\end{equation}
corresponding to the velocity field (in polar coordinates)
\begin{equation}
u_\theta =  \frac{K}{2\pi r} \;, \;\;\;\;\;\;\;\;
u_r = 0
\label{velpot}
\end{equation}
(The circulation $K$ is the line integral of the velocity over any
circuit including the origin; $\nabla \times \U = 0$ except at
the origin, where the velocity is evidently singular.)
We construct more complicated flows by superposing vortices at different
points ${\bf r}_j$.  (The vortex is an extended object; ${\bf r}_j$
denotes the position of the singularity.)
In a system of $N_V$ point vortices, 
each vortex $j$ moves in the velocity field defined by the 
superposition of all vortices except vortex $j$ 
itself \cite{sommerfeld}.  (For $N_V \leq 3$ the system is 
integrable \cite{leoncini}.)  Thus, in this rather special case
we can construct a complex fluid motion {\it without solving
the Euler equation}, by integrating the motion
of a system of $N$ point particles.
This makes the vortex system
particularly attractive for simulating incompressible, 
inviscid flow. 

Point-vortex systems have been used
for some time in studies of two-dimensional 
turbulence \cite{mccomb,aref,kraichnan} and of chaotic advection
\cite{aref84,leoncini}, and appear to be relevant to atmospheric dynamics  
on various scales \cite{dritschel}. 
Two interesting scaling properties of tracers in systems of four or more
point vortices are worth noting \cite{leoncini}: (1) the tracers exhibit anomalous
difusion, with the mean-square displacement growing $\sim t^{1.8}$; (2)
the lifetime $s$ of vortex pairs follows a power-law distribution,
$P(s) \sim s^{-2.7} $.  (Tracers are typically
excluded from the immediate vicinity of a vortex, but may on
the other hand become trapped at the periphery of a vortex pair.)
Compared with direct integration of
the Euler or Navier-Stokes equations, the computational demands
are orders of magnitude smaller.  Of course, one is restricted
to a two-dimensional, inviscid fluid.  (In the three-dimensional 
case the vortices become vortex lines, which stretch
and fold under the flow.  But such a system may still offer computational
advantages.)

In Ref. \cite{rain} I study a system of interacting point
vortices on the unit square with periodic boundaries.   
The velocity of vortex $i$ is given by
\begin{equation}
{\bf v}_i = \sum_{j \neq i} \frac{K_j}{2 \pi r_{ij}^2} 
\hat{\bf k} \times {\bf r}_{ij} ,
\label{vorvel}
\end{equation}
where $K_j$ represents the circulation of vortex $j$ 
(equal numbers of clockwise and anticlockwise vortices 
are used),
and {\bf r}$_{ij} = {\bf r}_i - {\bf r}_j$, under periodic boundaries, 
using the nearest-image criterion.  The velocity {\bf u}({\bf x},$t$)
at an arbitrary point {\bf x} in the plane (not occupied by a vortex)
is given by a similar sum including contributions from all vortices.
The number of vortices $N_V$ ranges from 10 to 126.

Several types of vortex-strength distributions are studied;
the simplest assigns all vortices the
same strength $|K|$.  
Other studies employ a hierarchical vortex
distribution, defined as follows.  The zeroth ``generation" consists
of a pair of vortices with $K = \pm K_0$.  Subsequent generations,
$n = 1,...,g$ have $2^{n+1}$ vortices, with circulation $|K| = K_0/\alpha^n$. 
I study $\alpha$ = 2, 3, and 4,  
using $g\!+\!1 = 5$ or 6 generations. 
The purpose of 
the hierarchical distribution is to provide structure on a variety of
length scales, without trying to reproduce any specific energy spectrum 
$E (k)$.  
The vortices are assigned random initial positions, 
but their subsequent evolution is deterministic \cite{note4}.  

Being point objects, the vortices possess no intrinsic length scale.
(Note however that in the presence of other vortices, the `sphere of
influence' of vortex $i$ is proportional to $K_i$.)
A characteristic length scale 
is the mean separation $\sim 1/\sqrt{N_V}$ between vortices.  
The vortex system defines a mean speed 
$u = \langle |{\bf u}({\bf x},t)| \rangle \propto K_0 \sqrt{N_V}$;
an important time scale is $\tau_C \sim 1/u$, the typical time for a fluid
particle to traverse the system.
A typical velocity field in a system of ten vortices (all of equal intensity)
is shown in Fig. 1.

A large number of
tracers, $N_p = 10\;000$, are thrown 
at random into a small region (a square of side 0.05), representing
a localized condensation event.  
(Alternatively, the tracer-laden region may be interpretated
as a parcel of atmosphere of high humidity, destined to
generate precipitation.)
In the analysis of rain and drought events, the observation 
interval $T$ plays an important role.  At time zero the vortices
begin their motion, and the tracers are inserted. 
The dynamics is followed up to
time $T$, when the simulation ends.

In the model, `rain' corresponds to the presence of one or more
tracers in a very small predefined region or `weather station',
of linear dimension 0.01.  At each
step of the integration, the number
of particles $n_i(t)$ at each station $i$ is monitored.  
A sequence of nonzero 
occupation numbers at a given station constitutes a rain event, just as
in the radar observations \cite{peters}; the intensity of a rain event is
$I = \sum_t n_i(t)$ where the sum is over the set of consecutive
time steps for which $n_i (t) > 0$.  In case $n_i =0$, station $i$ is
said to experience a drought.  The durations of droughts and of
rain events are likewise monitored over a time interval $T$. 

Fig. 2 shows successive configurations of a system of 10$^4$ particles and
10 vortices of equal intensity,
at times 0.46, 0.48 and 0.50, under conditions such that 
$\langle |{\bf u}({\bf x},t)| \rangle = 4$.  
(Thus $T=0.5$ corresponds to $2 \tau_C$.)
In this example the tracer-bearing region has become wrapped a vortex 
pair, and becomes increasingly stretched.
The tracers are
widely scattered, but their distribution remains highly nonuniform,
characterized by bands of high particle concentration.
(The tracer-free regions centered on the vortex pair
arise because the fluid trajectories circulate about the vortices,
so that tracers cannot penetrate this region from outside.)
At later times (see Fig. 3, for $T=8 \tau_C$)
tracers are more uniformly distributed,  
but there are again empty regions
centered on vortices or vortex pairs.   
(In these studies the tracers were released from a region of
size $0.01 \times 0.01$ to provide enhanced spatial resolution.)

Varying the vortex distribution and observation interval $T$,
the following trends emerge.  For $T/\tau_C$ in the range 0.1 - 2,
power-law rain-intensity and drought duration distributions
are found, as in Eqs. (\ref{PI}) and
(\ref{PD}).
The rain-intensity distribution follows a power law
over 4 - 5 1/2 decades, with an exponent 
$\tau_I$ in the range 0.93 - 1.02.
The drought-duration distribution
decays with a somewhat larger exponent, 1.12 - 1.16, and follows
a power law over 3 - 4 decades.  Larger exponent values are
associated with higher values of $\alpha$; these yield somewhat
smaller ranges for the power laws.  Conversely, the largest power-law 
range, and smallest exponent values, are observed when all vortices
are of equal strength.
There is no significant difference between 
the distributions obtained initially
and those found after the vortices have had some time to 
evolve, suggesting that the equilibration process 
expected in two-dimensional turbulence \cite{aref,kraichnan}
is not important as regards rain and drought statistics. 

Systems with varying numbers 
of vortices yield similar distributions, if we scale the intensity
$K \sim 1/\sqrt{N_V}$.  This is seen from the data collapse in
Fig. 4, in which results for systems of 10, 20, 50 and 100 vortices
(with $N_p=10^4$, $T \simeq 0.85 \tau_c$, and
$K=0.3$ for $N_V=10$), are shown.
Even systems with as few as ten vortices yield good power laws,
indicating that chaotic advection
is the essential feature leading to scale invariance,
rather than well developed turbulence.  
 
For larger values of $T/\tau_C$ the particles are more dispersed,
and the rain size and drought duration follow a stretched-exponential
form $P_i(I) \propto \exp(-C I^\beta)$ with $C$ a constant and
$\beta \simeq 0.5$.
Even for large values of $T/\tau_C$ (up to 200 in the present study), 
the distributions decay more slowly than an exponential, 
showing that the tracer density is non-Poissonian.   

The results of \cite{rain} may be summarized as
showing scale-invariant rain-size and drought-duration distributions
for intervals such that the tracers remain highly clustered.  
Although the decay exponents
are somewhat smaller than those obtained from observational data
(1.36 and 1.42 for rain size and drought duration, resp.
\cite{peters}), the simulations also show the drought duration 
decaying more rapidly than that for rain event sizes.  For conditions under
which the rain is more thoroughly dispersed, simulations yield 
stretched-exponential distributions.  
It is worth noting that the finding of non-power-law distributions 
at longer times does not signal an inability of the model to 
reproduce the observational results.  Rain, after all, does not
remain in the air indefinitely.  (It would, of course, be 
interesting to have
some way of comparing the model timescale $\tau_C$ with the typical
residence time of rain in the atmosphere.)  The tendency toward
a more uniform tracer distribution at times $\gg \tau_C$ is
in fact exagerated by the periodic boundaries of the model, and might
occur more slowly under a corresponding vortex dynamics in the atmosphere.

Even in a system as simple as that considered here, there is a large 
parameter space to be explored: number, circulation, and intensity of 
vortices, size and shape of the initial particle-bearing region, observation time $T$.
To close this section I report some preliminary results on 
situations not considered in \cite{rain}.  In all cases there are ten 
vortices, all of intensity 0.3, yielding 
$\langle |{\bf u}({\bf x},t)| \rangle = 4$.
A study in which the tracers are released from a circular, rather than
a square region yields the same exponents $\tau_I$ and $\tau_D$ as
found previously.  Thus the shape of the initial region appears not
to influence the event statistics.

It is natural to ask how relaxing the ``neutrality condition" (equal numbers
of vortices with clockwise and anticlockwise circulation) affects the
event distributions, since there is no obvious reason to assume 
such neutrality.
A study using all vortices with the same circulation again yields power-law
distributions, but with somewhat different exponent values, depending
on the observation time.
Specifically, 
for $T = 0.8 \tau_C$ I find $\tau_I = 1.01(1)$ and $\tau_D = 1.10(2)$,
similar to the results for the neutral system, while
for $T=1.2 \tau_C $, $\tau_I = 1.21(1)$ and
$\tau_D = 1.06(1)$.  Thus, allowing a net circulation
results in a larger rain intensity exponent at longer times, 
while the drought exponent is slightly smaller.  

There is also evidence that releasing 
the tracers from a smaller region (of linear size 0.01 instead of 0.05)
yields a larger $\tau_I$ and smaller $\tau_D$.  A study using $T=0.5$ (and equal numbers
of clockwise and anticlockwise vortices), yielded $\tau_I = 1.10(2)$, while
$\tau_D \simeq 1.02$.
Although it is difficult to draw firm 
conclusions from these preliminary results,
they demonstrate the generality of power-law distributions at
intermediate times, while
suggesting that exponent values may change depending
on the flow regime.

\section{Spatial Distribution of Tracers}

As discussed in the preceding section, a very simple model of passive tracers
in a velocity field defined by a system of point vortices is capable
of yielding power-law rain-intensity and drought-duration distributions
\cite{rain}.  
These results for events at a fixed observation site
suggest that the spatial arrangement of the tracers is somehow
related to the event distributions.  One might even hope to
understand the scale-invariant event distributions as arising from
a fractal tracer pattern as it sweeps over the observation site. 
In this section I present results on the spatial
distribution of the tracer particles, which can be thought of
as analogous to the distribution of rain over a region experiencing
storms.  The results are for systems with equal numbers of
clockwise and anticlockwise vortices, all of equal intensity $K$.

As a first step, I consider the occupancy histogram
$N(n)$ upon partitioning the system into a fine mesh; $N(n)$
denotes the number of elements with tracer occupancy $n$.
The simulation cell is divided into 10$^4$ square regions or boxes
of side 0.01, and the box-occupancy histogram determined after 
allowing the particle configuration evolve for a time $T$.  Recall
that initially, a small number of boxes (25 or so) will have high
occupancies, while the rest are empty.  If the particles tend toward
a uniform distribution, we should expect the histogram to approach a
Poisson distribution, with parameter 1 (there are 10$^4$ particles)
for large $T$.  The simulation results instead indicate a tendency
to form a power-law distribution at short times, followed by
a stretched-exponential form at longer times.  
Fig. 5 shows histograms
at various observation times for a system of 10 vortices of
equal intensity, $|K| = 0.3$.  At times $T/\tau_c =0.2$ and 0.4,
a peak near occupancy $n=400$ is evident, a remnant of the
initially compact distribution.  The histogram follows a power law
$N(n) \sim n^{-\epsilon}$, for $n \leq 200$ or so, with $\epsilon
= 0.54(1)$ for $T=0.2 \tau_C$.  As $T$ increases, the exponent $\epsilon$
becomes larger, and the histogram (on log scales) begins to
curve downward, signaling a faster than power-law decay.  For
$T=0.5$ the histogram is well described by a stretched exponential,
$N(n) \propto \exp[-\mbox{const.} \times x^\beta]$ with $\beta \simeq 1/7$.  
Thus the histogram remains non-Poissonian even for rather long times.
For a system of 100 vortices (with $K$ scaled to maintain the mean velocity
constant as discussed in Sec. V), the histogram is power law (with
$\epsilon = 0.67$) for $T=0.2 \tau_C$, and tends to a stretched exponential
(with $\beta \simeq 1/5$) for longer times.

In principle, the fractal dimension of the instantaneous particle 
distribution may be determined in a manner analogous to the fractal time
distribution described in Sec. III.  That is, we divide the system into 
ever-finer partitions (for example, squares of side $\ell=2^{-n}$ for 
$n = 1, 2, 3,...$) and determine the number $r(\ell)$ of occupied squares 
at scale $\ell$.  For uncorrelated positions we expect 
$r(\ell)  \propto \ell^{-2}$ away from the limits of very large or very
small boxes.  Applying this analysis to tracer configurations in the
vortex system yields $r(\ell)  \propto \ell^{-\gamma}$ with
$\gamma = 1.8$ - 1.9, depending on the observation time $T$ and
number of vortices.  This may signal an incipient fractal distribution,
but a glance at a typical configuration (Fig. 2) shows that
at the times of interest, the particle-filled region 
is not a fully developed
fractal structure, but rather is essentially linear, becoming
increasingly stretched (and wound about one or more vortices), 
and folded as times goes on.

The observation of a stretched, linear tracer-bearing
region suggests that we distinguish two directions, locally parallel
and perpendicular to the elongated region.  Observe that the particle
velocity is approximately parallel to the elongated direction.
Thus it is of interest to define coordinates $\eta$ and $\xi$ at any instant,
representing the distance from a given particle $i$ in directions 
parallel and perpendicular to (respectively) 
its velocity ${\bf v}_i$.
We study the tracer density as a function of distance from a
randomly chosen particle, along these directions, effectively
defining two-point correlation functions $C_{||}(x)$ and $C_\perp(x)$.
Fig. 6 shows that at time $2 \tau_C$
these functions are strongly peaked near the origin,
demonstrating a high degree of clustering, and that 
$C_{||}(x)$ is generally greater than $C_\perp(x)$, corresponding
to the elongated linear regions typical of the particle configuration
at intermediate times.   The correlation functions at time $8 \tau_C$
(shown in the inset of Fig. 6)
are much more uniform, away from the central peak, and appear to be
isotropic.

The configurations depicted in Fig. 2 suggest that
the repeated bands of particles (due to folding and/or wrapping
around a vortex) possess a nested structure.
We look for evidence of fractal structure along the perpendicular
direction $\xi$ by dividing this axis (along a narrow swath,
$|\eta| \leq 0.005$) into segments of length $\ell = 2^{-n}$, and 
determining the number of occupied segments $r(\ell)$ 
at scale $\ell$.  This function is shown for
various observation times in Fig. 7.  
At the shortest times $r(\ell)$ is constant for larger $\ell$,
indicating that only a single box is occupied at the larger scales,
due to the small insertion region (of length 0.01 here).
At intermediate times there is evidence of fractal scaling
(for example, at $T=2\tau_C$, $r \sim \ell^{-0.72}$ for
$\ell \leq 0.05$).  The slope $\gamma$ (away from
the saturation region at small $\ell$) appears to approach unity
at larger times, again signaling a more uniform tracer distribution.
(For $T=2 \tau_C$ the distribution $r(\ell)$ in the {\it parallel} 
direction is very similar to that for the perpendicular direction
shown in Fig. 7.)

The results for $r(\ell)$ cited above represent averages over
5 - 10 configurations.  High-resolution studies of single
configurations (involving 10$^5$
tracers released from a region of linear size 0.005), yield
power-law distributions in some cases, and stretched exponentials
in others, for the same parameter values.  For example,
a system with ten vortices ($|K|=1.2$, $T=4\tau_C$), one
realization yielded a stretched-exponential distribution
with $\beta \simeq 0.1$, while in other cases power laws
(with $\gamma = 0.45 - 0.55$, over three or more decades),
were found.  The stretched-exponential appears to be associated
with an overall scattering of tracers (as in Fig. 3) while in the
power-law case multi-band configurations predominate.  Similar
results are found in a system of 20 vortices.
Fig. 8 shows how the distribution evolves over time in a
typical high-resolution study. At short times,
$r (\ell) \sim 1/\ell$ for small $\ell$, indicating uncorrelated
positions, while at intermediate times and length scales there is evidence
of power-law scaling with $\gamma \approx 0.5$, and at longer times
the distribution can be fit to a stretched exponential with
$\beta \simeq 0.3$.

Summarizing the results described in this section, there is
preliminary evidence for a fractal tracer distribution at intermediate
times (on the order of $\tau_C$) associated with the nested filamentary
structures generated by stretching and folding of the particle-bearing
region.  The timescale for observation of a fractal tracer density
corresponds roughly to that associated with fractal rain and drought
distributions.  (One should recall, however, that the latter are 
accumulated from the time the tracers are released until
time $T$, whereas the tracer distributions discussed here are
instantaneous.)   It is easy to see that a fractal tracer distribution, 
swept past a fixed observation point, will generate 
power-law rain and drought distributions.  It remains to make
this connection more precise, a task complicated by the fact that
the characteristics of the tracer distribution
vary significantly over the observation period, and may also vary in space,
as a glance at Fig. 2 suggests.  This raises the possibility that the 
power-law distributions found in simulations and in actual measurements
represent a superposition of distributions associated with
different kinds of regions or events.  It would therefore be of interest
to identify simpler advection processes whose fractal properties 
can be determined with higher precision.

\section{Discussion}

I have reviewed observational evidence for fractal rain distributions,
and discussed a highly simplified model that points to chaotic
advection as the underlying reason.  The detailed properties
(e.g., the exponents associated with the power-law distributions)
furnished by the model differ from those found in observations.
(In truth, the present ``toy" model ignores so many important
atmospheric processes that quantitative agreement, if obtained, might well be
regarded as fortuitous.  The observational data, moreover, leave
doubt as to the universality of the power laws.)
Simulation of the model nevertheless leads to
the significant conclusion that neither interactions between
tracers (i.e., between raindrops or rain-bearing parcels of air),
nor fully developed turbulence are needed to generate power-law
rain and drought distributions.
The model also yields stretched-exponential event distributions
for longer observation times.
While the latter have not been reported, it is well to
recall that the observational data remain rather limited.  
Observations from other sites are needed to confirm the generality of 
power laws and the possibility of other (non-scale-invariant) 
forms.

Studying the occupancy statistics of boxes of various sizes, I find
evidence that the tracer density evolves,
under the vortex-system flow, to a fractal distribution 
at intermediate times.  The nature of this distribution, and its relation to the
power laws found for rain and drought events, needs to be studied in greater 
detail.

Clearly, the model employed in this proof-of-principle 
study contains a minimum of atmospheric
physics.  A three-dimensional description, 
allowing for stratification, convection, and vortex stretching would be 
desirable, as would inclusion of condensation, evaporation, and 
inertial effects \cite{benczik}.  
These improvements, all of which involve significant
computational complexity and expense, can be expected to alter detailed 
properties such as exponent values.  
The vortex model may readily be adapted to include some of these
effects, while others will require a full analysis of the coupled
Navier-Stokes and heat equations.

Since chaotic advection is an intrinsic
feature of atmospheric flow, one should expect
scale-invariant distributions to appear quite generally.  
In this regard it is interesting to note
that simulations of turbulent magnetohydrodynamic 
processes reproduce power-law burst 
distributions for solar flares \cite{boffetta,hughes}, and that tracer 
patterns similar to those reported
here are also found in simulations of two-dimensional barotropic 
turbulence \cite{haller}.  
Although (in the interest of simplicity) a closed model is analyzed
in this work, we should expect the same phenomenon to appear
in an open model with driving and dissipation \cite{aref}, due once again
to the chaotic nature of tracer motion.

In summary, I find that tracer distributions in 
two-dimensional flow, represented by a system of point
vortices, exhibit scale invariance during the early stage of the
dispersal process.
The event distributions are associated with fractal 
tracer distributions in space, produced by repeated stretching and folding
of fluid elemnts.
It therefore seems worthwhile 
to develop more realistic models, 
to understand the observations in greater detail.
Theoretical prediction of the rain and
drought distributions from
a model velocity field remains as a formidable challenge.
\vspace{1.5em}

\noindent {\bf {\small ACKNOWLEDGMENTS}}
\vspace{1em}

I thank Kim Christensen, Miguel A. Mu\~noz, Oscar N. Mesquita,
Ole Peters, Guilherme J. M. Garcia, Maya Paczuski and Francisco F. Araujo Jr. for helpful discussions.
This work was supported by CNPq, and CAPES, Brazil.

\newpage
\noindent FIGURE CAPTIONS
\vspace{1em}

\noindent FIG. 1. Velocity field in a system of ten vortices of equal strength.
\vspace{1em}

\noindent FIG. 2. Positions of 10$^4$ tracers (small points) and 10
vortices (open circles, clockwise circulation, filled, anticlockwise),
at times 0.46 (a), 0.48 (b) and 0.50 (c).
\vspace{1em}

\noindent FIG. 3. Positions of $5 \times 10^4$ tracers at time $8 \tau_C$, 
for the same conditions as Fig. 2.
\vspace{1em}

\noindent FIG. 4. Rain-size (main graph) and drought-duration (inset)
distributions in systems of vortices of equal strength,
$T \simeq 0.85 \tau_c$.  $\circ$: $N_V =10$; $\times$: $N_V=20$; 
$\Box$: $N_V=50$; $+$: $N_V=100$.  The vortex intensity $K$ is
scaled $ \sim 1/\sqrt{N_V} $ in these studies.
The straight lines have slopes of
-1.01 (rain size) and -1.13 (drought).
\vspace{1em}

\noindent FIG. 5. Instantaneous occupancy histogram $N(n)$ for 
boxes of side 0.01 in a system with 10 vortices, $|K|=0.3$. 
Filled squares: observation time $T=0.05$; $\Box$: $T=0.1$;
$\bullet$: $T=0.2$; $\circ$: $T=0.5$.
\vspace{1em}

\noindent FIG. 6. Correlation functions (unnormalized)
$C_{||}(x)$ ($\circ$) and $C_\perp(x)$ ($\bullet$) 
on semi-log scales, in
a system with ten vortices, observation time $T=2 \tau_C$.
Inset: a similar plot on linear scales for $T= 8 \tau_C$.
\vspace{1em}

\noindent FIG. 7. Distribution $r(\ell)$ along a line perpendicular to
the local velocity in a system of 10 vortices.
Observation times (bottom to top) $T/\tau_C =0.8$, 1.2, 2.0, 4.0,  
and 8.0.  (The data have been shifted vertically for visibility.)
\vspace{1em}

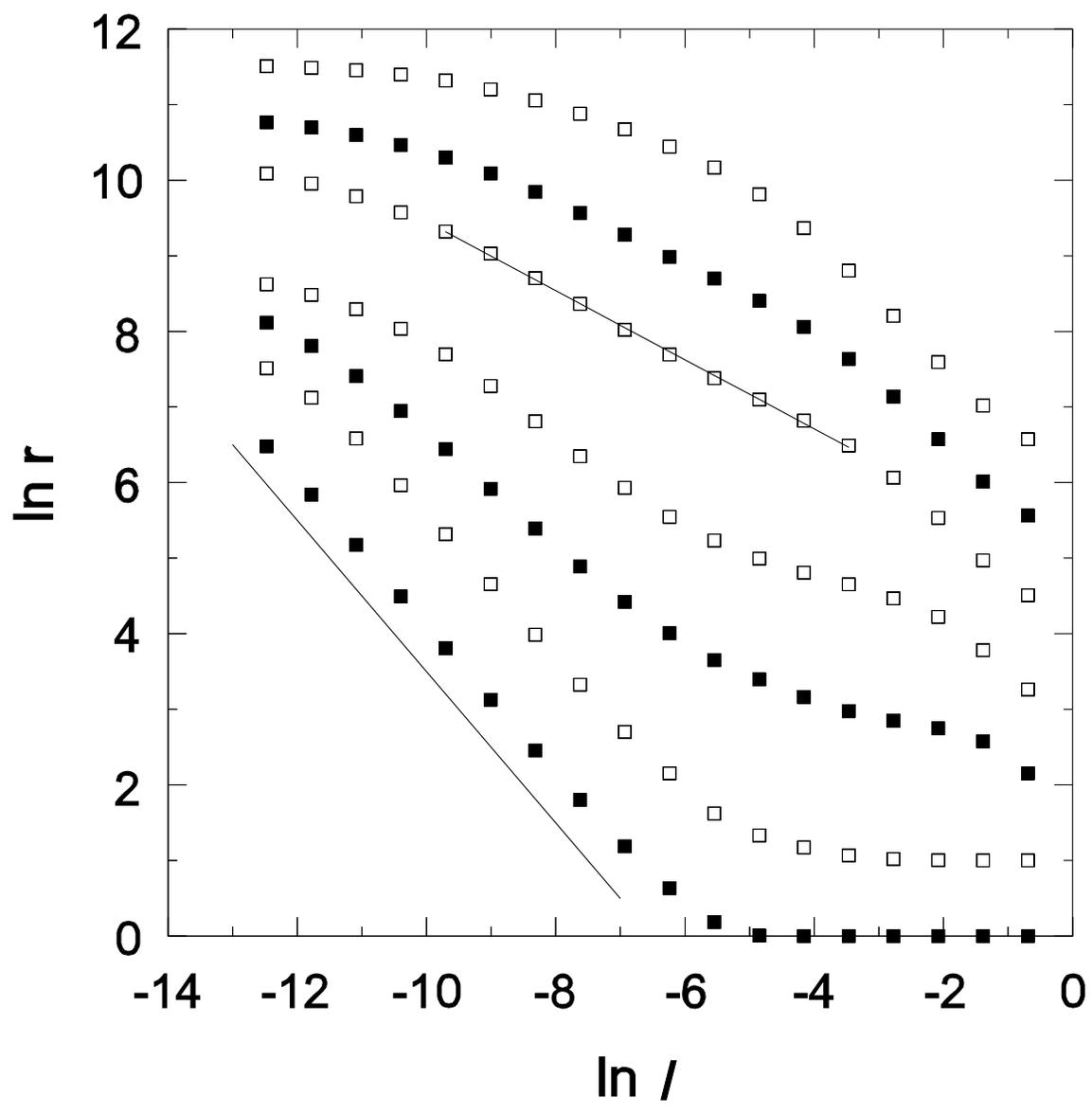
\noindent FIG. 8. Distribution $r(\ell)$ as in Fig. 7, but in a
single realization with $10^5$ tracers released from a region of size 0.005.
Observation times (bottom to top) $T/\tau_C =2.4$, 3.2, 4.0, 4.8,  
5.6, 6.4 and 8.0.  (The data have been shifted vertically for visibility.)
The slopes of the straight lines are -1 and -0.47.


\begin{references}
\bibitem{note1}
	 The literature on fractal rain distributions is
	 extensive; for a bibliography see \cite{lovejoy03}.
\bibitem{note3} 
	 The incompressibility condition should be well satisfied since the
	 velocities of interest are small compared to the speed of sound.
\bibitem{note5}
	 Further details and extensive references on systems of
	 point vortices may be found in Ref. \cite{leoncini}.
\bibitem{note4}
	 It would be of interest to study other initial
	 configurations, for example, one approximating a vortex sheet,
	 which should then suffer a Kelvin-Helmholtz instability.
	 The periodic boundaries might also be removed, allowing
	 the vortex system to attain its natural, unconstrained
	 size.
\end{references}
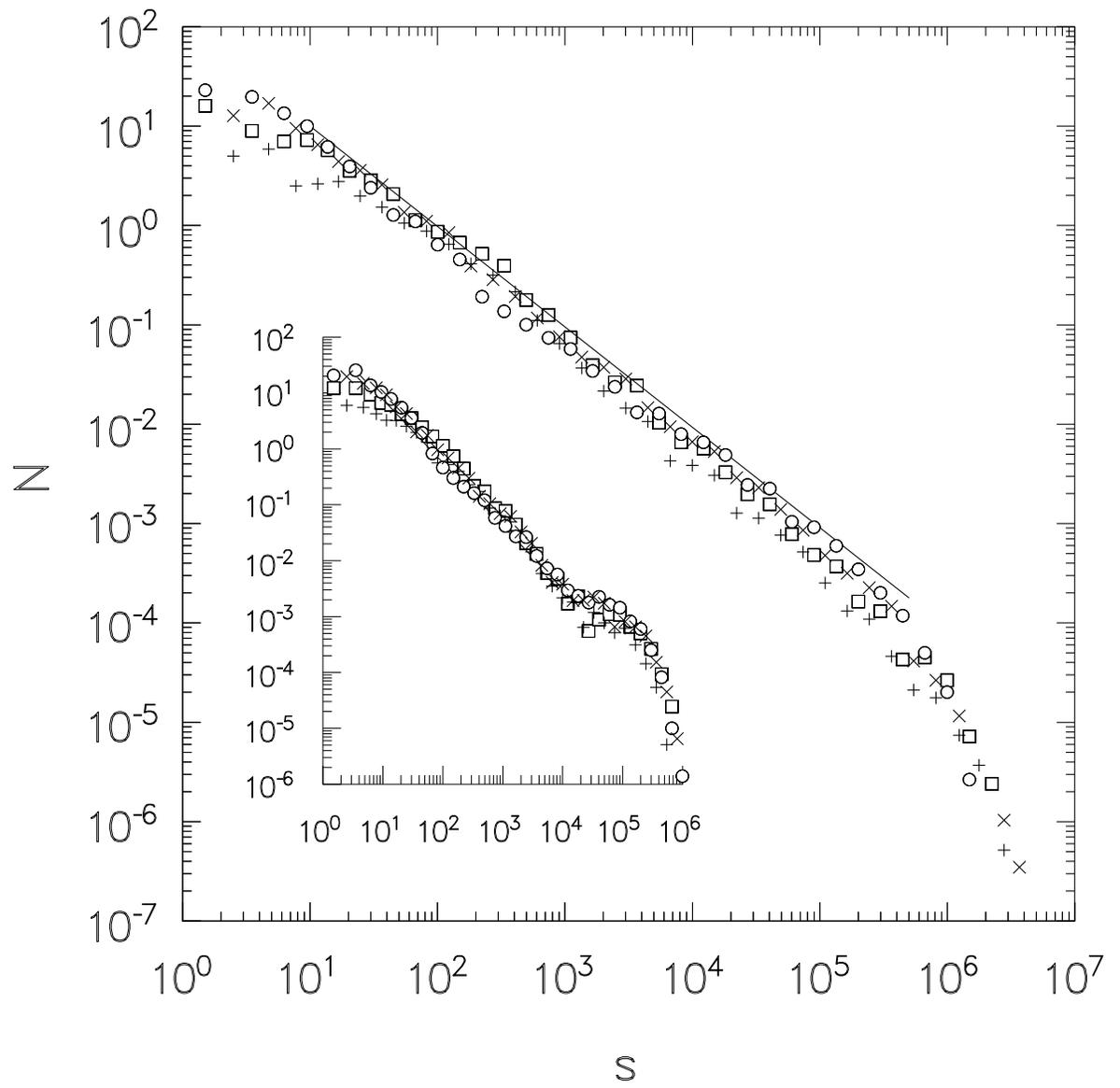
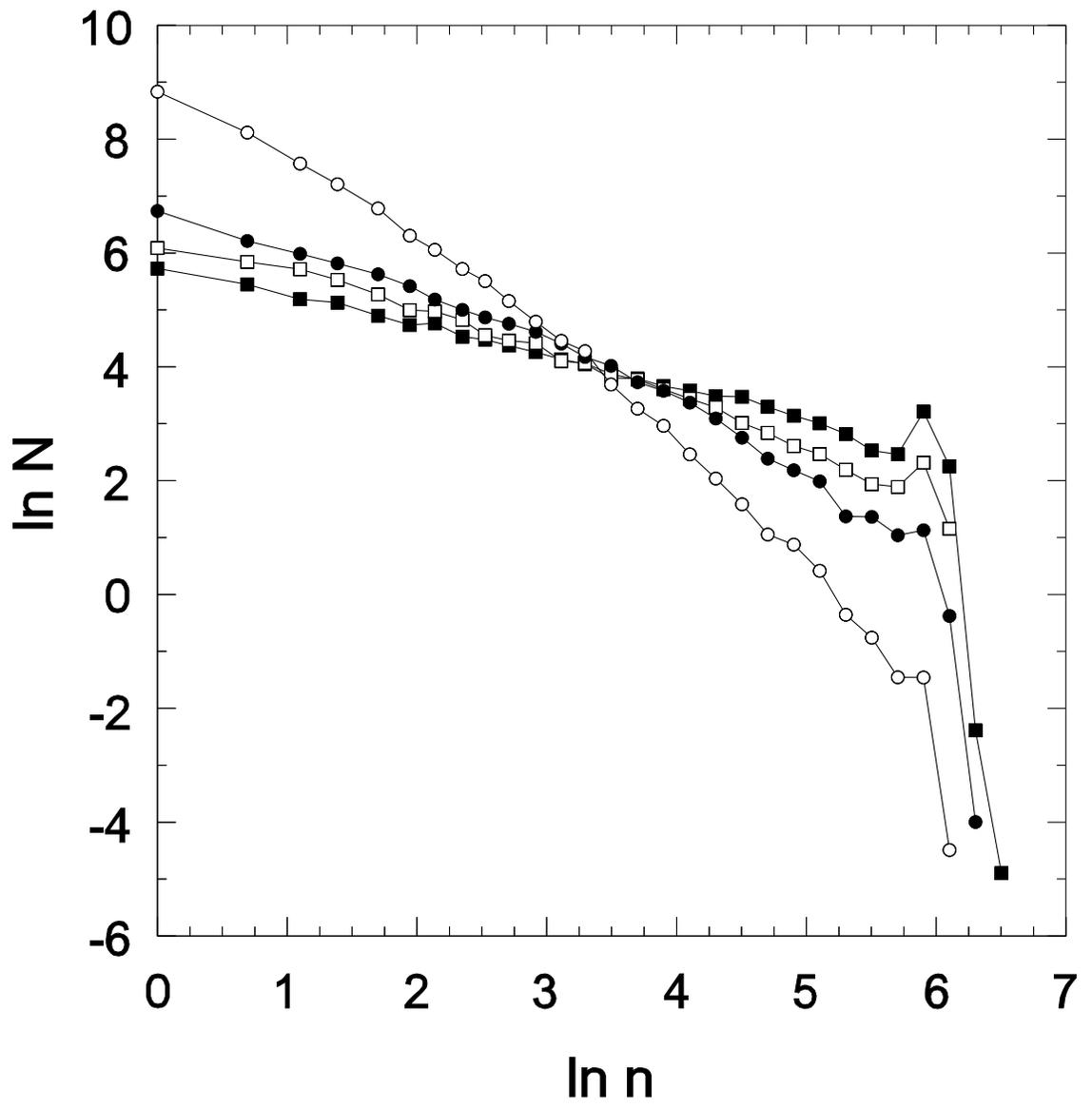
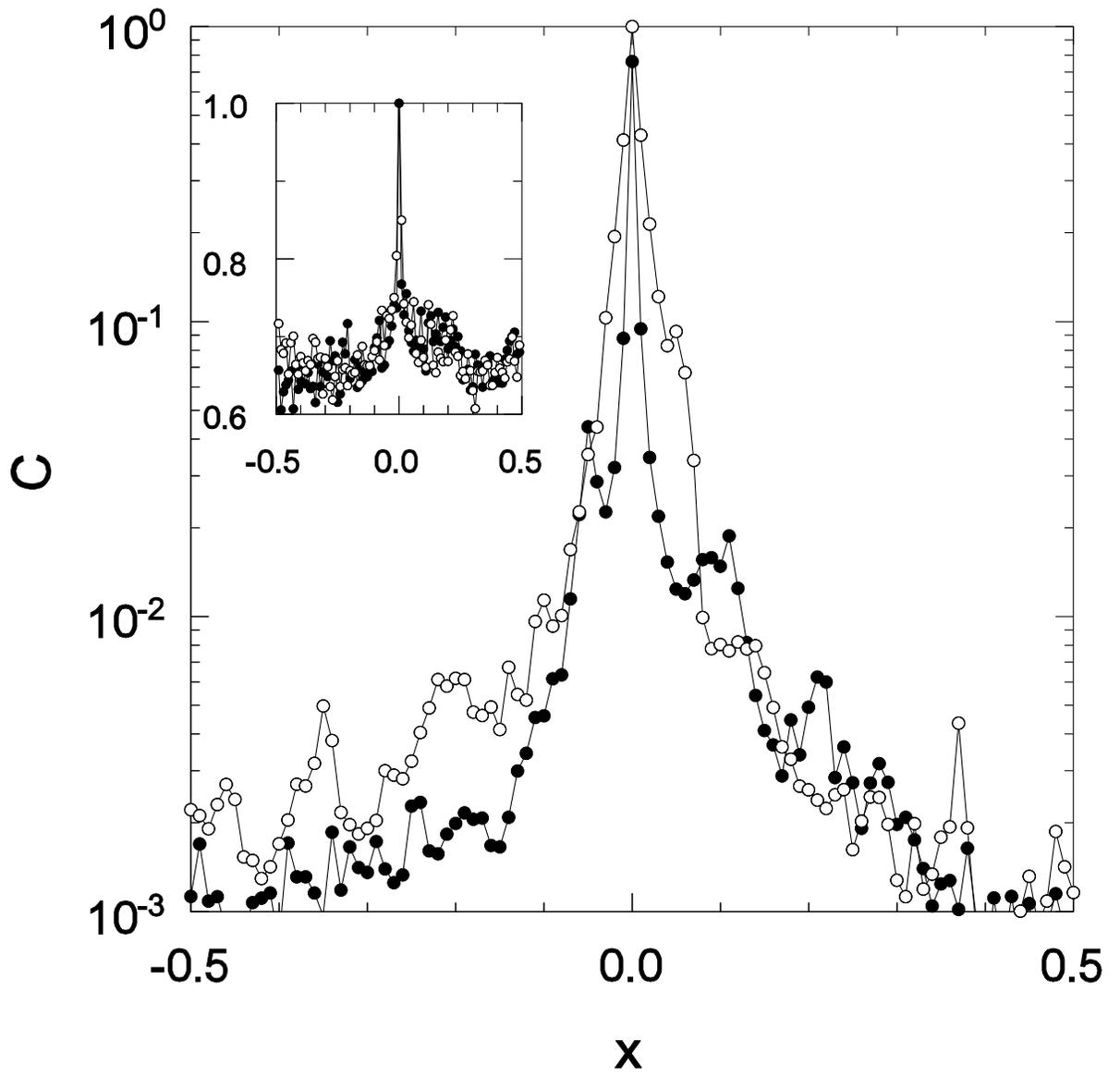
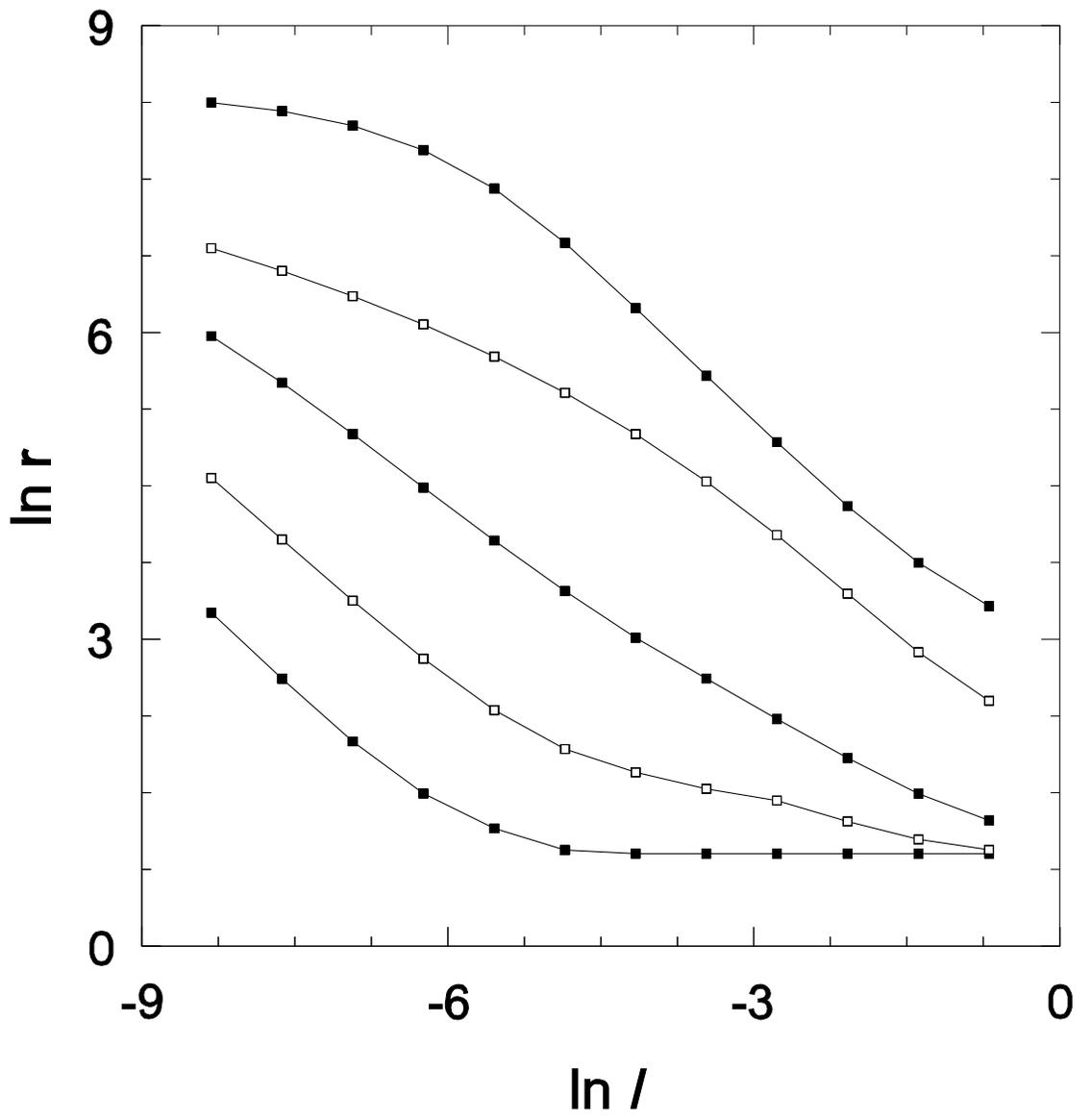
\end{document}